# Fast LDPC GPU Decoder for Cloud RAN

Jonathan Ling and Paul Cautereels

*Abstract*— The GPU as a digital signal processing accelerator for cloud RAN is investigated. A new design for a 5G NR low density parity check code decoder running on a GPU is presented. The algorithm is flexibly adaptable to GPU architecture to achieve high resource utilization as well as low latency. It improves over an existing layered design that processes additional codewords in parallel to increase utilization. In comparison to a decoder implemented on a FPGA (757K gate), the new GPU (24 core) decoder has 3× higher throughput. The GPU decoder exhibits 3 to 5× lower decoding power efficiency, as typical of a general-purpose processor. Thus, GPUs may find application as cloud accelerators where rapid deployment and flexibility are prioritized over decoding power efficiency.

## I. INTRODUCTION

Cloud radio access networks (RAN) facilitate the rapid development and deployment of new features while reducing the overall cost of ownership [1][2]. This is enabled by network virtualization, resource pooling, and lower cost development due to common-off-the-shelf hardware. Graphical processing units (GPUs) are typical in cloud computing and are used as accelerators for many applications. GPUs are characterized by high floating-point computation rate and high memory bandwidth to feed those execution cores. Field programmable gate arrays (FPGAs) are also candidates for programmable cloud acceleration. A comparative study [3] found that certain algorithms run faster and more cost effectively on the GPU.

The upper layers of the RAN stack are currently implemented on general purpose processors. An open issue is the effectiveness of cloud computing with GPU acceleration for the lower layers. One of the salient requirements of 5G is low latency. This puts stringent demands on algorithm design as hardware resources are limited and predefined. In this study we focus on the acceleration of 5G NR forward error correction low density parity check code (LDPC) [4] as this function consumes nearly half of the total uplink processing cycles.

Clearly GPUs excel at graphics acceleration and machine learning, but their application to digital signal processing isn't firmly established. Digital signal processing is often performed in fixed point to minimize cost and power and on smaller word sizes, e.g. 8 bits vs. typical 32 bits. On Nvidia GPUs small word sizes are supported by specialized single-instruction-multiple-data (SIMD) "video" instructions that operate on packed 32-bit word, processing up to four 8-bit instructions in parallel. These video instructions were utilized by researchers to develop a high throughput decoder [5]. Recent GPU architecture has changed such that these instructions are now emulated by up to 10 instructions, and in fact some key primitives required by the decoder are not provided. An interesting tradeoff is support for vector processing of two packed 16-bit floating point quantities.

Another challenge is related to the threads needed to hide pipeline latency. For example, 256 threads on the Pascal architecture and 128 threads on Volta/Tesla/Ampere architecture are necessary to hide basic arithmetic latency in a single streaming-multiprocessor (SM) core. This minimum level of parallelism must be expressible in the algorithm otherwise there will be resource underutilization.

A related challenge is to fit an algorithm's working set into local memory. Coarse grained parallelization of independent jobs increases memory and bandwidth requirements. The L1 and L2 caches in the GPU are small and access to main memory is on order of hundreds of cycles. Depending on the shared memory requirements, algorithms may need be re-worked to obtain additional parallelism, applying more threads to the same set of data, rather than simply increasing independent workload.

In this letter we describe techniques needed to reduce the latency of the forward error correction decoder for a 5G NR cloud base station. We flexibly exploit additional parallelism to improve the standard layered decoder. Fixed-point SIMD primitives to improve accelerator performance are described. Latency, throughput, power, and power efficiency are investigated and compared to an FPGA design [6].

The remainder of this letter is organized as follows: In Section II we provide the context of this study: a 5G system model for cloud RAN and the uplink processing chain. In Section III, we review belief propagation decoding and its implementations. In Section IV we describe our innovations to GPU LDPC decoding. In Section V performance measurements are presented and discussed. The summary and final remarks are found in the Conclusion.

## II. 5G SYSTEM MODEL

Figure 1 illustrates the architecture of a typical cloud RAN deployment [7]. The gNB is split into two components the CU and DU. The CU contains the upper SDAP and PDCP layers, while the DU contains the lower RLC, MAC, and PHY layers. A second low split point in the DU has been defined [8]. ORAN has specified a 7.2 split while a 7.3 split with its lower fronthaul requirements is also prevalent.

Timing requirements for uplink processing in 4G are determined by the hybrid-automatic-repeat-request (HARQ) process. An ACK/NACK from an FDD uplink reception must be send 4 ms later. This leaves 2 ms for uplink processing. Whereas for 5G preferably decoding should be completed within one slot, which may take values between 250 μS to 1 ms according to the parameter μ.

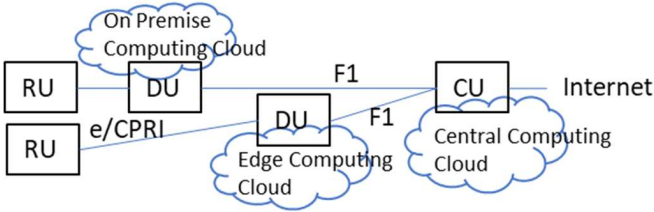

Figure 1 Cloud RAN architecture & placement.

3GPP has defined quasi-cyclic irregular LDPC codes for data channels. Two base graphs are specified, that is BG1 and BG2. Various length codewords are achieved through the lifting factor $Z$, which defines the submatrix size in the parity check matrix $H_{BG}$ [9]. The number of code word information bits is $K=Z \cdot K_b$, where $K_b$ is the number of information base graph columns. The code rate, defined as the ratio of information bits to the total number of code bits is $R=K/N_c$. $R$ controls the error correcting capability, and different rates are achieved by specifying a certain number of basegraph rows. For a base graph with $M_{bg}$ rows, then the total number of coded bits $N_c=Z(K_b+M_{bg})$. To improve performance the first two information columns are punctured, i.e. not transmitted, and are fully recovered by decoding [4]. Rows within the submatrix $Z_m$ are independent and can be performed in parallel.

## III. BELIEF PROPAGATION DECODING

LDPC codes may be efficiently decoded by a soft-decision belief propagation algorithm. The Tanner graph is a useful representation where an edge from the variable node to check node corresponds to a "1" in the parity check matrix H, as shown in Figure 2. The number of edges from a variable node corresponds to the column weight $w_c$, i.e. number of 1s in a column. Whereas the number of edges entering a check node corresponding to the row weight $w_r$, i.e. the number of 1s in a row. The Tanner graph also illustrates the decoding processes as the "messages" from variable-to-check nodes and check-to-variable nodes may be interpreted as probabilities. The check node equation produces the belief of a variable node given all the other input variable nodes. Let $i$ take values from 1 to $w_r(r)$ to index the set of columns involved in row $r$. Processing each at the check nodes is given by

$$L_{c \to v}^i = \sum \boxplus L_{v \to c}^j \quad \forall j \neq i$$
$$= 2\tanh^{-1}\left(\prod \tanh\left(\frac{L_{v \to c}^j}{2}\right)\right).$$

where $\boxplus$ denotes the xor operation. The min-sum approximation [10] simplifies the xor sum to

$$L_{c \to v}^i \approx \beta \min(|L_{v \to c}^j|) \prod \text{sgn}(L_{v \to c}^j) \quad \forall j \neq i$$

where $\beta$ is a correction factor. Let index $k$ index a set of rows involved in the column $v$. The variable node log-likelihood-ratio (LLR) is:

$$L_v = L_b + \sum_{k=1}^{w_c(v)} L_{c \to v}^k$$

where $L_b$ is the bit LLR obtained from the symbol-to-bit demapper. The next iterations' variable-to-check messages are found by:

$$L_{v \to c}^i = L_v^i - L_{c \to v}^i .$$

Iterations continue until a valid codeword is found or the maximum number of iterations is reached. After decoding the CRC check verifies the message correctness.

The flooding belief propagation decoder evaluates all parity check equations in parallel despite their dependencies. Alternatively, the horizontal layered decoder processes parity check equations one row at a time, feeding results from one dependent row to the next. As compared to the flooding decoder, memory requirements are reduced, and the number of iterations is approximately halved.

The parallelism in the flooding and that provided by the GPU appeared to be a natural fit [11]. Despite the apparent lack of parallelism, the layered decoder was found to be much faster than the flooding decoder on a GPU [5], probably due to the reduced memory required. Applying this result, we focus our investigation on the implementation a layered decoder for 5G cloud RAN.

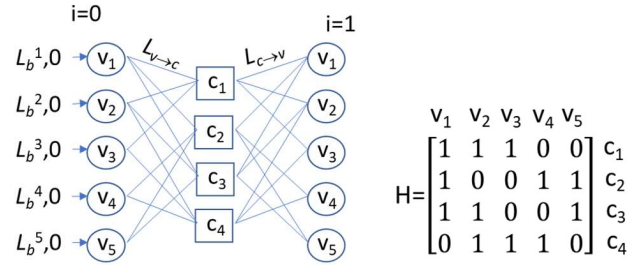

Figure 2. Tanner graph decoder structure for a LDPC code with parity check matrix H.

## IV. GPU IMPLEMENTATION STRATEGIES

The primary source of parallelism in the 5G NR LDPC is specified by the lifting factor $Z$. From Section III, it may be observed that the layered processing executes on one row at a time, with amount of work proportional to $w_r(r)$. It appears that the algorithm is sequential so that "exploitation of intraframe computation parallelism is quite impossible" [5]. Accordingly, this design now assigns one row per thread, and each thread works on $w_r(r)$ columns. However, by SIMD instructions the GPU can process 16-bit floating point, or 8-bit fixed point yielding $\rho=2$ or $\rho=4$ rows in parallel respectively. The *high-throughput* strategy thus engages

$$N_{thread} = \rho^{-1} Z$$

threads. Assuming a single codeword is processed per SM, $\rho=4$, then $1 \leq N_{thread} \leq 96$. which is clearly much less than optimal (at best 75% of the minimum on Turing architecture).

The storage size of $L_v$ may be calculated as $S_v = \varepsilon N_c$ (bytes) where $\varepsilon$ is the number of bytes per LLR. For 8-bit LLRs and BG1 $S_v^{MAX}$ is 26 kB per codeword. For 16-bit LLRs $S_v^{MAX}$ is 52 kB bytes. Storage for $L_{c \to v}$ may be calculated as $S_{c \to v} = \varepsilon Z \sum_{r=1}^{M_{bg}} w_c(r)$ (bytes). $S_{c \to v}^{MAX}$ is 121 kB for 8-bit LLRs, and 242 kB for 16-bit LLRs. For the larger code words both $L_v$ and $L_{c \to v}$ cannot fit in shared memory (64 kbytes Turing). One strategy is to keep $L_v$ in shared memory while placing $L_{c \to v}$ in main memory. The messages are needed less frequently and can be preloaded.

Attempting to improve utilization by processing an additional codeword doubles shared memory usage. $L_v$ might fit into shared memory but now reading and writing the messages $L_{c \to v}$ to main memory, i.e. memory bandwidth, becomes the limiting factor. Due to either the shared memory size and/or memory bandwidth, the high-throughput strategy cannot be used to increase utilization. This motivates us to search for additional avenues for parallelism.

We observe that that column processing is fully independent except for finding the min, and submin, and sign across all columns. These non-parallelizable steps can be performed partially in parallel by a technique called parallel reduce, which performs an $n$ step algorithm in $\log_2(n)$ steps. The GPU has a programming paradigm called cooperative groups and special instructions to efficiently shuffle data between the threads within a warp. For BG1 the available parallelization varies between 3 and 19, but the number of threads $\alpha$ in a cooperative group but must be factor of 2 for full efficiency. This means that $\alpha$ threads processes up to $\lceil w_r(r)/\alpha \rceil$ columns. Each thread processes $\rho$ rows at a time, and the number of columns per thread may not be uniform. Therefore, a *low-latency* strategy engages a total of

$$N_{thread} = (\alpha/\rho)Z$$

threads. $\alpha$ can be adjusted according to $Z$, to keep utilization high. This strategy flexibly breaks up work over a cooperative thread group to reduce the latency of a single codeword. In contrast the high-throughput strategy only decodes additional codewords with additional threads. The high-throughput strategy yields fixed latency regardless of code size $Z$. Whereas latency increases with $Z$ for the low-latency strategy. This may be especially helpful in decoding short packets for 5G ultra-low latency applications.

In more detail, first, up to $\lceil w_r(r)/\alpha \rceil$ elements of the LLRs $L_v$ and previous $L_{c \to v}$ are loaded from memory. $L_{v \to c}$ is computed. Four intermediate quantities: min, sub-min, message sign, and variable sign, are computed. These are defined as:

$$m_1 = \min(|L_{v \to c}^i|) \quad \forall i$$
$$m_2 = \min(|L_{v \to c}^i|) \quad \forall i \neq \arg\min(|L_{v \to c}|)$$
$$s_{v \to c} = \prod \text{sign}(L_{v \to c}^i) \quad \forall i$$
$$s_v = \prod \text{sign}(L_v^i) \quad \forall i$$

where $i$ take values from 1 to $w_r(r)$ to index the set of columns C involved in row $r$. They are computed first per thread, and then across threads by parallel reduce in $\log_2(\alpha)$ steps. Figure 3 illustrates for block of $\alpha=4$ cooperating threads, which start with intermediate values, and to return the same final value to all threads.

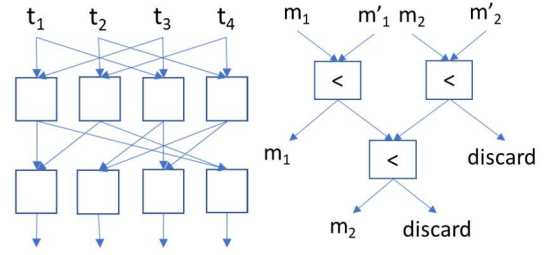

Figure 3. Parallel reduce and $m_1$ $m_2$ calculation block.

During the reduce phase, inside the operations block of Figure 3, $m_1$ and $m_2$ are computed by two ordering and a single selection operation. An ordering intrinsic was devised to minimize cycle count. For 8-bit LLRs packed into 32-bit word, the pseudo "C" code is:

```
[min, max ] = ordVec8(uint32_t a, uint32_t b)

    uint32_t mask = __vcmpltu4( a, b)
    min = ( mask & a) | (~mask & b)
    max = (~mask & a) | ( mask & b)
```

where vcmplte4u is a CUDA intrinsic that does vector comparison using 4 assembly instructions. The last two lines are translated into single LOP3 assembly instruction each. Thus ordVec8 takes 6 assembly instructions. For half precision 16-bit LLRs packed into a 32-bit word the pseudo "C" code is:

```
[min, max ] = ordVec16(uint32_t a, uint32_t b)

    uint32_t ans
    asm( "{set.le.f16x2.f16x2 %0,%1,%2;\n}"
      : "=r" (ans) : "r" (a) , "r" (b) );
    ans = ans << 2
    asm ("prmt.b32 %0,%1,0,0xbb99;" :
      "=r" (mask) : "r" (ans))
    min = (mask & a) | (~mask & b)
    max = (~mask & a) | (mask & b)
```

where set.le.f16x2.f16x2 performs the vectorized comparison and prmt.b32 generates a mask from the upper word bits. ordVec16 consists of 1 floating point instruction and 4 integer instructions. Thus, ten instructions, i.e. +67%, are needed to produce the same number of results as ordVec8.

## V. PERFORMANCE RESULTS

The performance of the low-latency (LL) GPU decoder with $\alpha$ is set to 4 is compared to a FPGA decoder [6] given metrics of latency, throughput, power and decoding power efficiency. The FPGA decoder was built on a moderate sized FPGA with 747K logic cells, consuming ½ the area. The layered GPU decoder was tested on a consumer grade mid-sized GTX1660 with TU116-400 chip with an observed a boost clock speed of 1995 MHz and a total of 24 SMs.

### A. Latency and Throughput

The latency of the GPU and FPGA decoders are compared in Table 1, along with a flooding GPU decoder [6]. All GPU decoders exhibit much higher latency. The GPU while a vector processor, still incus excess latency that is typical of general-purpose processors due to the need to move intermediate data between registers vs. the customized pipeline of the FPGA. As shown in Section IV, the principal operation in finding $m_1$ and $m_2$, is a SIMD ordering function that requires multiple assembly instructions. Nonetheless the low-latency decoders meet our eMBB decoding target of 500 µS at 10 iterations. The fact that the flooding GPU decoder has 30× latency of LL-INT highlights both the importance and difficulties in tailoring an algorithm to the GPU's architecture.

Throughput is given in Table 2, and the situation changes dramatically where the LL-INT GPU decoder is about 3× the throughput of the FPGA. To obtain the FPGA throughput, we assume full utilization of logic area and 20 iterations. On the GPU all SMs were engaged running the integer decoder whereas only half of the SMs could be engaged running the float decoder. This is because the messages need to be loaded from external memory, and at exactly ½ the number of SMs the chip runs out of bandwidth. Attempting further loading of SMs increases latency substantially. The processing rate is about the same for other code rates such as R=1/3.

|  | FPGA Flooding | GPU Flooding | GPU LL-INT | GPU LL-FP |
|---|---|---|---|---|
| BG1 R=1/3 | 8.5 µS* | 1040* µS | 35 µS | 40 µS |
| BG2 R=1/5 | 6.2 µS* | 660 µS* | 27 µS | 30 µS |

Table 1. Maximum latency for different decoders per iteration (* per two iterations).

|  | FPGA | GPU LL-INT | GPU LL-FP |
|---|---|---|---|
| BG1 R=1/2 | 0.6 Gcbs | 1.8 Gcbps | .9 Gcbs |

Table 2. Throughput in giga-coded-bps per chip.

*B. Power Efficiency*

GPU power measurements were taken using the nvidia-smi utility. There is high base power of about 40 watts, and much smaller incremental power per SM. Interestingly when LL-INT decoder runs completely in shared memory, per SM power is 70% lower. Combining the power measurements with the throughput to obtain power efficiency yields 80 Mcb/J for FPGA, and 27 to 15 Mcb/J for GPU. Thus, the FPGA is 3× to 5× more power efficient than the GPU.

|  | FPGA (W/instance) | GPU (W/SM) |
|---|---|---|
| Z=384 R=1/3 | 7.5 | 3.5 + 1.6 base |
| Z=128 R=1/3 | - | 1.2 + 1.6 base |

Table 3: Measured power for FPGA and GPU decoders.

## VI. CONCLUSION

A new low latency design strategy has been presented that exploits additional parallelism in the decoding process. Our design solves the problem of underutilization of resources while maintaining the same memory working set. Overall throughput is quite good as compared to an FPGA implementation. We note that GPU architecture is continuously being improved. Instruction set changes as well as increases in cache and memory bandwidth will continue to improve throughput and power efficiency. Yet, programmable and fixed accelerators will retain their power efficiency. Most likely the cloud RAN will contain a mixture of accelerators, as economics and deployment flexibility dictate.